\documentstyle[12pt]{article}

\font\tenmsb=msbm10 scaled\magstep 1
\font\sevenmsb=msbm7 scaled \magstep 1
\font\faivemsb=msbm5 scaled \magstep 1
\newfam\msbfam
\textfont\msbfam=\tenmsb
\scriptfont\msbfam=\sevenmsb
\scriptscriptfont\msbfam=\faivemsb\def\Bbb#1{{\fam\msbfam #1}}

\font\tengothic=eufm10 scaled\magstep 1
\font\sevengothic=eufm7 scaled\magstep 1
\newfam\gothicfam
\textfont\gothicfam=\tengothic
\scriptfont\gothicfam=\sevengothic

\newcommand{\be}{\begin{equation}}
\newcommand{\ee}{\end{equation}}
\newcommand{\dlt}{\delta}
\newcommand{\br}{{\bf r}}
\newcommand{\ba}{{\bf a}}
\newcommand{\cD}{{\cal D}}
\newcommand{\cA}{{\cal A}}
\newcommand{\cX}{{\cal X}}
\newcommand{\ra}{\rightarrow}
\newcommand{\cH}{{\cal H}}
\newcommand{\cL}{{\cal L}}
\newcommand{\vp}{\varphi}
\newcommand{\cY}{{\cal Y}}
\newcommand{\bt}{\beta}
\newcommand{\al}{\alpha}
\newcommand{\prt}{\partial}
\newcommand{\dgr}{\dagger}
\newcommand{\Om}{\Omega}

\begin{document}

\begin{center}

{\Large{\bf Mesoscopic Phase Fluctuations: General Phenomenon in
Condensed Matter} \\ [5mm]
V.I. Yukalov} \\ [3mm]
{\it Bogolubov Laboratory of Theoretical Physics \\
Joint Institute for Nuclear Research, Dubna 141980, Russia}

\end{center}

\vskip 1cm

\begin{abstract}

General conditions for the occurrence of mesoscopic phase fluctuations
in condensed matter are considered. The description of different
thermodynamic phases, which coexist as a mixture of mesoscopically
separated regions, is based on the {\it theory of heterophase fluctuations}.
The spaces of states, typical of the related phases, are characterized by
{\it weighted Hilbert spaces}. Several models illustrate the main features
of heterophase condensed matter.

\end{abstract}

\vskip 1cm


\vskip 2cm

{\parindent=0pt

}

\newpage

\section{Introduction}

In many cases, condensed matter does not consist of a unique thermodynamic
phase but is separated into mesoscopic regions filled by different phases.
A number of examples of such matter are described in review [1]. The
coexisting phases can be distinguished by their order parameters that change
from one mesoscopic region to another. The order parameters are often
associated with some special symmetry [2]. However, this is not compulsory,
and in many cases different phases cannot be characterized by different
kinds of symmetry. For instance, crystalline and fluid phases have different
spatial symmetry. Magnetic and paramagnetic phases possess different magnetic
symmetry, which can be observed by means of neutron scattering [3]. Similarly,
ferroelectric and paraelectric phases possess different symmetry characterized
by polarization. If the latter phases coexist, the polarization of the mixed
system is spatially nonuniform [4,5]. Contrary to this, liquid and gas cannot
be distinguished by their symmetry, though they may have different densities
or other properties [6]. Metal and nonmetal are also distinguished not by
symmetry but by their conductivity. Better and worse conducting mesoscopic
regions can coexist in many materials [7--9].

{\it Mesoscopic} is a key word here, implying the following. Suppose that
inside one thermodynamic phase there appear the nuclei of another phase.
Each nucleus, in order to represent a coherent germ of a phase, should have
a characteristic coherence length $l_{coh}$ much longer than the mean
interparticle distance $a$. As is evident, an object of the size $a$ cannot
be associated with a thermodynamic phase. At the same time, the coherence
length is to be essentially shorter than the characteristic length $L$ of the
whole system. Thus, the phase germs are of the mesoscopic size [1], that is
$$
a \ll l_{coh} \ll L \; .
$$
This property is to be understood on the average. Generally, the shape of
heterophase germs is not necessarily regular, but can be rather complicated.
Several germs can also touch each other forming very ramified fractal-type
objects, similar to fractal animals at criticality [10]. The total effective
length of such a fractal formation can reach the length $L$ of the whole
system. Being mesoscopic in size, the heterophase nuclei are randomly
distributed in space, forming no regular structures.

{\it Random} is another key word. Heterophase germs are random in space,
analogously to random disordering defects in solids [11,12], but being
different from the latter by the mesoscopic size of the phase germs.

These germs, being mesoscopic in space, can also be mesoscopic in time.
This means that they may be not static but dynamical, arising and then
disappearing, with a characteristic lifetime, or coherence time $t_{coh}$,
such that
$$
\tau_{loc} \ll t_{coh} \ll T_{obs} \; ,
$$
where $\tau_{loc}$ is the time of local equilibrium and $T_{obs}$ is
an observation time [1]. The temporal behaviour of these fluctuational
dynamic germs is chaotic. When the system is chaotic in time and random in
space, it is termed turbulent. Since here the spatio-temporal chaoticity
is connected with the dynamics of heterophase nuclei, this phenomenon may
be called {\it heterophase turbulence} [13]. The origin of heterophase
turbulence can be classified into three types, extrinsic, intrinsic, and
stochastic [13].

{\it Extrinsic} origin of turbulence assumes the existence of sufficiently
strong external forces acting on the system. Such forces can induce
chaoticity in classical and quasiclassical systems [14,15].

{\it Intrinsic} origin presupposes that the system behaves chaotically
owing to its own internal properties. There are many examples of such
chaotic dynamical systems [16,17]. For statistical systems this means that
quasiequilibrium heterophase states are their {\it chaotic attractors}
[1,13,18].

{\it Stochastic} origin of heterophase turbulence is intermediate between
the above two. It takes into account that all real systems are never
completely isolated but can be only {\it quasi-isolated}, and that an
infinitesimally small external stochastic perturbation can trigger chaoticity,
making the system {\it stochastically unstable} [1,18,19].

The most interesting cases are those related to the intrinsic and stochastic
origins, when heterophase states are not forced by external sources but are
self-organized. There exist plenty of examples of such systems, reviewed in
Ref. 1. Among those heterophase materials that have attracted much attention
in recent days, it is possible to mention liquid crystals [20],
high-temperature superconductors [21--24] and colossal magnetoresistant
materials [25]. Mesoscopic phase fluctuations lead to the occurrence of many
unusual features of matter, for instance such as colossal magnetoresistence
[25] and anomalous diffusion in solids [26]. Their appearance may transform
a second-order phase transition into a first-order one [1]. They can induce
different anomalies in the behaviour of thermodynamic and dynamic
characteristics of matter, for instance: the maxima of specific heat below
the critical temperature [27], specific-heat jumps at the nucleation point
[1,28], anomalous saggings of the M\"ossbauer factor at the phase transition
point [29,30], anomalous decrease of sound velocity and increase of isothermal
compressibility at the point of a structural transition [29,30], and other
anomalies [1].

All unusual features of condensed matter with mesoscopic phase fluctuations
can be self-consistently described in the frame of the statistical {\it
theory of heterophase fluctuations} [1]. In order not to force readers to
study the whole review [1], a brief survey of this theory is presented in
Sections 2 and 3. The main purpose of this presentation is, from one side,
to give a rather short account of the principal ideas, the theory is based
on, and from another side, to clearly point at the principal mathematical
techniques providing the foundation. An attempt of combining brevity with
accuracy in outlining the theory is a methodical novelty of the paper.
Several examples of sections 4 to 6 illustrate applications of the theory.
The generality of the approach to treating mesoscopic phase fluctuations
as well as the generality of the phenomenon itself are emphasized in
Section 7.

\section{Description of Phase Separation}

When talking about phase separation, one, first of all, keeps in mind the
separation of thermodynamic phases in real space. Different phases can be
distinguished by their different properties. Those phase characteristics
that are the most appropriate for distinguishing different phases are called
the order parameters. The order parameter of a thermodynamic phase is a set
$\eta$ of quantities typical of the phase and uniquely distinguishing it from
other phases. If there are several phases, each of them is characterized by
the corresponding order parameter $\eta_\nu$, with the index $\nu=1,2,\ldots$
enumerating the phases. When a physical system, occupying a real-space
volume $\Bbb{V}$, is phase separated, then this volume can be divided into
subvolumes $\Bbb{V}_\nu$ filled by different phases with different order
parameters $\eta_\nu$. A family $\{\Bbb{V}_\nu\}$ of subvolumes $\Bbb{V}_\nu$
forms an {\it orthogonal covering} of $\Bbb{V}$, such that
\be
\label{1}
\Bbb{V} = \bigcup_\nu \Bbb{V}_\nu \; , \qquad
\Bbb{V}_\mu \bigcap \Bbb{V}_\nu =\dlt_{\mu\nu}\; \Bbb{V}_\nu \; .
\ee
The latter can be represented as the resolution of unity
$$
\sum_\nu\xi_\nu(\br) = 1 \qquad (\br \in \Bbb{V})
$$
for the {\it manifold indicator functions}
\begin{eqnarray}
\label{2}
\xi_\nu(\br) =\left \{ \begin{array}{ll}
1, & \br \in \Bbb{V}_\nu \\
0, & \br\not\in\Bbb{V}_\nu \; .
\end{array} \right.
\end{eqnarray}
A family of indicator functions
\be
\label{3}
\xi \equiv \{ \xi_\nu(\br) | \; \nu=1,2,\ldots; \; \br\in\Bbb{V} \}
\ee
uniquely defines a {\it phase configuration} in the real-space volume
$\Bbb{V}$.

In physical systems, the separation of thermodynamic phases is usually
realized so that these phases are not sharply split apart, but are
transformed one into another through an interphase layer, where the order
parameter continuously changes from the value typical of one phase to
that characteristic of another phase. How could we then locate a separating
surface between such phases? First, the location of the separating surface
can be fixed in a unique manner by employing the Gibbs method of separating
surfaces [31]. This can be done by invoking some additional conditions. For
instance, the {\it equimolecular separating surface} is uniquely defined
by requiring that the surface density of matter be zero. Moreover, a precise
location of the separating surface is not important. This is because the
phase configuration in what follows will be varied, with the family of
indicator functions (3) being treated as a random variable.

In order to define a method of varying phase configurations, we should
consider a manifold $\{\xi\}$ of all possible families (3) and to rig this
manifold with a differential measure. To this end, we introduce an orthogonal
subcovering $\{\Bbb{V}_{\nu i}\}$ of the volume $\Bbb{V}_\nu$, so that
\be
\label{4}
\Bbb{V}_\nu = \bigcup_{i=1}^{n_\nu} \Bbb{V}_{\nu i} \; , \qquad
\Bbb{V}_{\mu i} \bigcap \Bbb{V}_{\nu j} = \dlt_{\mu\nu} \; \dlt_{ij}
\Bbb{V}_{\nu i} \; .
\ee
To each subregion $\Bbb{V}_{\nu i}$, we ascribe a vector
$\ba_{\nu i}\in\Bbb{V}_{\nu i}$ playing the role of a local center of
coordinates, so that moving $\ba_{\nu i}$ implies a congruent motion of
$\Bbb{V}_{\nu i}$. The shapes of the subvolumes $\Bbb{V}_{\nu i}$ as well
as the location of the centers $\ba_{\nu i}$ inside each $\Bbb{V}_{\nu i}$
can be arbitrary. For a given subcovering (4), the indicator functions (2)
can be decomposed as
\be
\label{5}
\xi_\nu(\br) = \sum_{i=1}^{n_\nu} \xi_{\nu i} (\br -\ba_{\nu i}) \; ,
\ee
where
\begin{eqnarray}
\xi_{\nu i} (\br-\ba_{\nu i}) = \left\{ \begin{array}{ll}
1, & \br \in \Bbb{V}_{\nu i} \\
\nonumber
0, & \br \not\in \Bbb{V}_{\nu i} \; .
\end{array} \right.
\end{eqnarray}
The differential measure on the manifold $\{\xi\}$ is introduced [1] as the
differential functional measure
\be
\label{6}
\cD\xi \equiv \lim_{\{ n_\nu\ra\infty\} } \; \dlt\left ( \sum_\nu x_\nu -1
\right ) \prod_\nu dx_\nu \; \prod_\nu \prod_{i=1}^{n_\nu}
\frac{d\ba_{\nu i}}{V} \; ,
\ee
in which
$$
x_\nu \equiv \frac{1}{V} \; \int_{\Bbb{V}} \xi_\nu(\br)\; d\br \; ,
\qquad V \equiv \int_{\Bbb{V}} \; d\br \; .
$$
The measure (6), with varying $x_\nu\in[0,1]$ and $\ba_{\nu i}\in\Bbb{V}$,
induces a topology on the manifold $\{\xi\}$. This results in the topological
{\it configuration space}
\be
\label{7}
\cX \equiv \{ \xi|\; \cD\xi \} \; ,
\ee
composed of all possible phase configurations.

There exists another principal problem connected with phase separation. This
is how such a macroscopic quantity as a thermodynamic phase is related to the
microscopic states of a physical system. For generality, we consider this
question for quantum systems.

Suppose that all admissible microscopic quantum states of a physical system
form a Hilbert space $\cH$. Let $\{\vp_n\}$ be an orthonormal basis in this
space, such that $(\vp_m,\vp_n)=\dlt_{mn}$. Then the {\it space of microstates}
can be presented as a closed linear envelope
\be
\label{8}
\cH \equiv \overline\cL\{ \vp_n\}
\ee
over the basis $\{\vp_n\}$. Any state $f\in\cH$ can be written as a
decomposition
$$
f = \sum_n f_n\vp_n \; , \qquad f_n \equiv (\vp_n,f) \; .
$$
For any two states $f,f'\in\cH$, their scalar product is given by the
equation
$$
(f,f') = \sum_n f_n^* f_n' \; .
$$

Now assume that for each representative $\vp_n$ of the basis $\{\vp_n\}$ a
weight $p_\nu(\vp_n)$ is assigned, normalized so that
$$
\sum_n p_\nu(\vp_n) = 1 \; .
$$
This defines the {\it weighted basis} $\{\vp_n,p_\nu(\vp_n)\}$. The {\it
weighted scalar product} is introduced as
\be
\label{9}
(f,f')_\nu \equiv \sum_n p_\nu(\vp_n)f_n^* f_n' \; .
\ee
Then a closed linear envelope over the weighted basis results [1] in
the {\it weighted Hilbert space}
\be
\label{10}
\cH_\nu \equiv \overline\cL\{ \vp_n,p_\nu(\vp_n)\} \; ,
\ee
with the associated scalar product (9).

From definition (9), one has
\be
\label{11}
(\vp_m,\vp_n)_\nu = p_\nu(\vp_n)\dlt_{mn} \; , \qquad
(\vp_n,f)_\nu = p_\nu(\vp_n) (\vp_n,f) \; .
\ee
For an operator $A_\nu$, given on $\cH_\nu$, the trace operator is defined
as
\be
\label{12}
{\rm Tr}_{\cH_\nu}\; A_\nu \equiv \sum_n (\vp_n,A_\nu\vp_n)_\nu \; .
\ee
From equalities (11), it follows that
$$
(\vp_m,A_\nu\vp_n)_\nu = p_\nu(\vp_m)(\vp_m,A_\nu\vp_n) \; .
$$
Therefore the trace (12) reads
\be
\label{13}
{\rm Tr}_{\cH_\nu}\; A_\nu \equiv \sum_n p_\nu(\vp_n)(\vp_n,A_\nu\vp_n) \; .
\ee
Using the above formulas, we may specify the {\it weighted average}
\be
\label{14}
<A_\nu> \; \equiv {\rm Tr}_{\cH_\nu}\; \hat\rho_\nu \; A_\nu
\ee
of an operator $A_\nu$, with $\hat\rho_\nu$ being a statistical operator on
$\cH_\nu$, normalized so that
$$
{\rm Tr}_{\cH_\nu}\; \hat\rho_\nu = 1 \; .
$$

The order parameters $\eta_\nu$ can be presented as the averages
\be
\label{15}
\eta_\nu \equiv \; <\hat\eta_\nu>
\ee
of the corresponding order operators $\hat\eta_\nu$. Each order parameter
is assumed to possess the properties classifying the related thermodynamic
phase. These properties implicitly determine the appropriate basis weight
$p_\nu(\vp_n)$. The latter is not uniquely defined by Eq. (15), but,
generally, there can exist a class of such basis weights satisfying the same
 Eq. (15). Any representative of this class is acceptable, provided all of
them yield the coinciding averages. In practice, it is even not necessary
to find an explicit form of $p_\nu(\vp_n)$, but it is sufficient to impose
the appropriate restrictions on the averages, which would guarantee the
typical features of the related phases [1]. There are several ways of
selecting pure phases, such as the Bogolubov method of infinitesimal
sources [32], the method of thermodynamic quasiaverages [1,33], the method
of order indices [34,35], and others [1]. A very convenient way is by
imposing additional conditions on density matrices or Green functions [36].
A similar way of imposing conditions on distributions or correlation
functions is also suitable for classical systems.

\section{Heterophase Statistical State}

The total space of microstates for a system with phase separation is the
weighted fiber space
\be
\label{16}
\cY \equiv \otimes_\nu \cH_\nu \; ,
\ee
whose fibering yields the weighted Hilbert spaces (10).

Let $A_\nu(\xi_\nu)$ be an operator on $\cH_\nu$, when a phase configuration
(3) is fixed. One may also consider the operator density $A_\nu(\xi_\nu,\br)$
for which
\be
\label{17}
A_\nu(\xi_\nu) = \int A_\nu(\xi_\nu,\br)\; d\br \; .
\ee
Here and in what follows, the integration with respect to $\br$, if not
specified, implies the integration over the whole system volume $\Bbb{V}$.
The operator density on the total space (16) is defined as
\be
\label{18}
A(\xi,\br) \equiv \oplus_\nu A_\nu(\xi_\nu,\br) \; ,
\ee
so that the related operator is
\be
\label{19}
A(\xi) \equiv \int A(\xi,\br)\; d\br = \oplus_\nu A_\nu(\xi_\nu) \; .
\ee
Examples of operator densities are the Hamiltonian density $H(\xi,\br)$ and
the number-of-particle operator density $N(\xi,\br)$.

Statistical properties of a heterophase physical system, with a fixed phase
configuration, can be characterized by the quasiequilibrium Gibbs ensemble [1].
A particular phase configuration $\xi$ is connected with the local inverse
temperature $\bt(\xi,\br)$ and local chemical potential $\mu(\xi,\br)$. The
latter enter in the {\it quasi-Hamiltonian}
\be
\label{20}
Q(\xi) \equiv \int \bt(\xi,\br) \left [ H(\xi,\br) - \mu(\xi,\br)
N(\xi,\br)\right ] \; d\br \; .
\ee
This defines the quasiequilibrium thermodynamic potential, which will be called
{\it quasipotential}
\be
\label{21}
q \equiv -\; \frac{1}{N} \; \ln\; {\rm Tr}_\cY \; \int
e^{-Q(\xi)}\; \cD\xi \; ,
\ee
with $N$ being the total number of particles and where the integration with
respect to $\xi$ is over the configuration space (7) of all possible phase
configurations. The quasiequiblibrium statistical operator is
\be
\label{22}
\hat\rho(\xi) \equiv
\frac{\exp\{-Q(\xi)\}}{{\rm Tr}_\cY \int\exp\{-Q(\xi)\}\cD\xi} \; .
\ee
Expectation values of operators $A(\xi)$ are given by the averages
\be
\label{23}
<\tilde A> \; \equiv {\rm Tr}_\cY \;
\int \hat\rho(\xi)A(\xi)\; \cD\xi \; .
\ee

Integrating with respect to phase configurations, we assume that measuring
procedures effectively average the observable quantities over random phase
configurations. If all parts of the heterophase system ate locally equivalent
to each other, then the local inverse temperature and local chemical potential,
being averaged over phase configurations, should not depend on $\br$, which can
be formulated as the {\it condition of heterophase equilibrium}
\be
\label{24}
\bt = \int \bt(\xi,\br)\; \cD\xi \; , \qquad
\mu = \int \mu(\xi,\br) \; \cD\xi \; .
\ee

Accomplishing the integration over the configuration space (7), we keep in
mind the existence of the thermodynamic limit
$$
N\ra \infty \; , \qquad V\ra \infty \; , \qquad
\frac{N}{V} \ra const \; .
$$
Then the thermodynamic quasipotential (21) reduces [1] to
\be
\label{25}
q = -\; \frac{1}{N}\; \ln{\rm Tr}_\cY \; e^{-\bt\tilde H} \; ,
\ee
with the effective Hamiltonian
\be
\label{26}
\tilde H = \oplus_\nu H_\nu \; , \qquad H_\nu = H_\nu(w_\nu) \; .
\ee
The terms
$$
H_\nu = \int \left [ H_\nu(w_\nu,\br) - \mu N_\nu(w_\nu,\br) \right ]\;
d\br
$$
of the Hamiltonian (26) can be called [1] the {\it phase-replica Hamiltonians}.
The operator densities $A_\nu(w_\nu,\br)$ have the same form as
$A_\nu(\xi_\nu,\br)$, but with $\xi_\nu$ replaced by $w_\nu$. The quantities
$w_\nu$ are the {\it phase probabilities}, playing the role of average geometric
probabilities for the corresponding phases. These probabilities are defined [1]
as the minimizers of the thermodynamic quasipotential (25), that is,
\be
\label{27}
q ={\rm abs}\min_{ \{ w_\nu\} } q(\{ w_\nu\} ) \; ,
\ee
under the condition
\be
\label{28}
\sum_\nu w_\nu =1 \; , \qquad 0 \leq w_\nu \leq 1 \; .
\ee
For the operator averages (23), we find [1]
\be
\label{29}
<\tilde A> \; = {\rm Tr}_\cY \; \hat\rho\; \tilde A \; ,
\ee
where the effective statistical operator
\be
\label{30}
\hat\rho \equiv
\frac{\exp(-\bt\tilde H)}{{\rm Tr}_\cY\exp(-\bt\tilde H)} \; ,
\ee
is expressed through the Hamiltonian (26). The operator $\tilde A$ in the
right-hand side of Eq. (29) is
\be
\label{31}
\tilde A \equiv \oplus_\nu A_\nu \; , \qquad A_\nu =  A_\nu(w_\nu) \; .
\ee
From Eqs. (26) and (30) it follows that
\be
\label{32}
\hat\rho = \otimes_\nu \hat\rho_\nu \; , \qquad \hat\rho_\nu \equiv
\frac{\exp(-\bt H_\nu)}{{\rm Tr}_{\cH_\nu}\exp(-\bt H_\nu)} \; .
\ee
Therefore the average (29) can be written as
\be
\label{33}
<\tilde A> \; = \sum_\nu \; < A_\nu> \; .
\ee
The set $\{\tilde A\}$ of all operators, corresponding to observable
quantities, composes an algebra of local observables $\tilde \cA$. The {\it
heterophase statistical state} is the family $<\tilde \cA>$.

The operators from the algebra of local observables can be divided into
single-particle, two-particle, and, generally, $k$-particle operators, so
that the overall structure of an operator $A_\nu$ is
\be
\label{34}
A_\nu = \sum_k \frac{1}{k!}\; w_\nu^k \; A_\nu^{(k)} \; .
\ee
The standard Hamiltonians of statistical systems usually contain the
single-particle and the two-particle terms only. Hence the structure of a
replica Hamiltonian is
\be
\label{35}
H_\nu = w_\nu \; H_\nu^{(1)} + \frac{1}{2}\; w_\nu^2\; H_\nu^{(2)} \; .
\ee
Assuming that $H_\nu^{(k)}$ do not depend on $w_\nu$, we may explicitly
derive the conditions of heterophase stability.

Consider, for concreteness, a heterophase mixture of two phases. It is
convenient to introduce the notation
\be
\label{36}
w \equiv w_1 \; , \qquad w_2 = 1 - w\; .
\ee
Then the phase probability $w$ is a minimizer of the thermodynamic
quasipotential $q=q(w)$, which implies the equations
\be
\label{37}
\frac{\prt q}{\prt w} = 0 \; , \qquad \frac{\prt^2 q}{\prt w^2} > 0 \; .
\ee
The first of the latter, taking into account form (25), yields
\be
\label{38}
<\frac{\prt\tilde H}{\prt w} > \; =  0 \; .
\ee
The second of Eqs. (37) has the meaning of the positive definiteness of the
{\it heterophase susceptibility}
\be
\label{39}
\tilde\chi \equiv \frac{\prt^2 q}{\prt w^2} > 0 \; .
\ee
Taking account of Eq. (25), we have
\be
\label{40}
\tilde\chi = \frac{\bt}{N}\left [ <\frac{\prt^2\tilde H}{\prt w^2} > \; - \;
\bt <\left ( \frac{\prt\tilde H}{\prt w}\right )^2 > \right ] \; .
\ee
Then the stability condition (39) reads
\be
\label{41}
\left ( <\frac{\prt^2\tilde H}{\prt w^2} >\right ) > \; \; \bt <
\left ( \frac{\prt\tilde H}{\prt w}\right )^2> \; .
\ee
Introducing the averages
\be
\label{42}
K_\nu \equiv \frac{1}{N} \; < H_\nu^{(1)} > \; , \qquad
\Phi_\nu \equiv \frac{1}{N} \; < H_\nu^{(2)}> \; ,
\ee
we find
$$
\frac{1}{N}\; < \frac{\prt\tilde H}{\prt w}> \; K_1 + w\Phi_1 -
K_2 -(1-w)\Phi_2 \; , \qquad
\frac{1}{N}\; < \frac{\prt^2\tilde H}{\prt w^2} > \; = \Phi_1 +\Phi_2 \; .
$$
The extremum condition (38) yields an equation for the phase probability
\be
\label{43}
w = \frac{\Phi_2+K_2-K_1}{\Phi_1+\Phi_2} \; .
\ee
From the inequalities $0\leq w\leq 1$ it follows that
\be
\label{44}
-\Phi_1 \leq K_1 - K_2 \leq \Phi_2 .
\ee
And the stability condition (41) gives
\be
\label{45}
\Phi_1 + \Phi_2 > \frac{\bt}{N} \left [ < \left (
\frac{\prt\tilde H}{\prt w}\right )^2 > \right ] > 0 \; .
\ee
Inequalities (44) and (45) can be called the conditions of {\it heterophase
stability}. These are to be complimented by the usual conditions of
thermodynamic stability, such as the positivity of specific heat and of
isothermal compressibility.

\section{Theory of Melting and Crystallization}

An important application of the developed approach is to describing the phase
transitions of melting and crystallization [37,38]. The mesoscopic fluctuations
inside a crystal are the liquid-like droplets presented by the regions of local
disorder of the crystalline lattice, arising below melting temperature. Above
the latter, mesoscopic heterophase fluctuations are quasicrystalline clusters
inside liquid.

To consider this phase transition, let us take the replica Hamiltonian in the
standard form
$$
H_\nu = w_\nu \int \psi_\nu^\dgr(\br) \left ( -\; \frac{\hbar^2\nabla^2}{2m_0}
- \mu\right ) \psi_\nu(\br)\; d\br \; +
$$
\be
\label{46}
+ \frac{1}{2}\; w_\nu^2 \int \psi_\nu^\dgr (\br) \psi_\nu^\dgr(\br')
\Phi(\br -\br') \psi_\nu(\br')\psi_\nu(\br) \; d\br\; d\br' \; ,
\ee
in which $\psi_\nu(\br)$ are field operators. The chemical potential $\mu$ can
be defined from the equation
\be
\label{47}
N = N_1 + N_2 \; , \qquad N_\nu \equiv -\; < \frac{\prt H_\nu}{\prt\mu} > \; .
\ee
Here, we deal with a two-phase mixture, because of which $\nu=1,2$.

The local density of particles in a $\nu$-phase is
\be
\label{48}
\rho_\nu(\br) \equiv \; <\psi_\nu^\dgr(\br)\psi_\nu(\br)> \; .
\ee
The number of particles in the $\nu$-phase, according to the definition in
Eq. (47), becomes
\be
\label{49}
N_\nu = w_\nu \int \rho_\nu(\br)\; d\br \; .
\ee
The average density of the $\nu$-phase is
\be
\label{50}
\rho_\nu = \frac{1}{V} \int \rho_\nu(\br) \; d\br \; .
\ee
And the mean density of particles in the whole system writes
\be
\label{51}
\rho \equiv \frac{N}{V} =  w_1\rho_1 + w_2\rho_2 \; .
\ee
If the density $\rho$ is fixed, then the latter equation defines the chemical
potential $\mu=\mu(\rho,T)$ as a function of density and temperature.

Let the crystalline phase be numbered with $\nu=1$, while the liquid phase,
with $\nu=2$. By this assumption, the local density (48) for the crystalline
phase is periodic,
\be
\label{52}
\rho_1(\br+\ba) = \rho_a(\br) \neq const \; ,
\ee
where $\ba$ is any lattice vector. But the liquid phase has a uniform density
\be
\label{53}
\rho_2(\br) = \rho_2(0) = const \; .
\ee
Since it is the local density that distinguishes the phases, $\rho_\nu(\br)$
can be treated as the order parameter. Alternatively, it is possible to define
the order parameter
\be
\label{54}
\eta_\nu \equiv \sup_\br \; \frac{\rho_\nu(\br)}{\rho_\nu} - 1 \; .
\ee
For an ideal crystal, $\sup_\br\rho_1(\br)=\rho_1(\ba)$, but for a liquid,
$\rho_2(\br)=const$. Therefore
\be
\label{55}
\eta_1 = \frac{\rho_1(\ba)}{\rho_1} - 1 > 0 \; , \qquad \eta_2 \equiv 0 \; .
\ee

The analysis [37,38] of the typical behaviour of the phase probabilities
$w_\nu$ results in the following figures. In Fig. 1, the probability of the
crystalline phase is shown as a function of temperature $\Theta$ in arbitrary
units. At zero temperature, the system is in pure crystalline phase. For higher
temperatures, the crystalline-phase probability slightly diminishes along the
line {\bf AB} and then abruptly falls down from the point {\bf B} to {\bf E}
at the melting temperature $\Theta_m$. Above the latter, the curve {\bf EF}
corresponds to the probability of crystalline clusters inside the liquid phase.
The branches {\bf BC} and {\bf DE} are related to metastable states of an
overheated crystal and crystalline clusters inside an overcooled liquid,
respectively.

Figure 2 displays the probability of the liquid phase $w_2=1-w_1$. At zero
temperature, $w_2=0$. With increasing temperature, there appear a small
probability of liquid-like disordered regions in the crystalline phase, which
rises along the line {\bf ab}. At the melting temperature $\Theta_m$, the
probability of the liquid phase jumps up to the point {\bf e}, after which
it follows the line {\bf ef}. The branches {\bf bc} and {\bf de} again are
related to metastable states.

In the similar way, one may treat a liquid-glass transition. A glassy state
can be described by a disordered lattice. Then the local density $\rho_1(\br)$
is not periodic, but, at the same time, it is not constant. The glassy solid
state and liquid state can be distinguished by the order parameter (54), so
that, similarly to Eqs. (55), one has
\be
\label{56}
\eta_1 > 0 \; , \qquad \eta_2 = 0 \; .
\ee
Possible behaviour of the glassy-phase probability is shown in Fig. 3.
Contrary to the crystal-liquid transition, the glass-liquid transition can
occur as a continuous transition, without the jumps of phase probabilities
at the glassyfication temperature $\Theta_g$. The glassy state is metastable,
being not an absolute extremum of a thermodynamic potential, as compared to
the crystalline state.

Thermodynamic potentials for heterophase matter are the same as for pure-phase
systems. We can consider, for instance, the grand potential
$$
\Omega = -pV = \Omega(T,V,\mu) \; ,
$$
the free energy
$$
F=\Omega +\mu N = F(T,V,N) \; ,
$$
or the Kramers potential
$$
K = -\; \frac{\Omega}{T} = K(\bt,V,\bt\mu) \; .
$$
All of them can be expressed through the quasipotential (25) by means of the
equations for the grand potential
\be
\label{57}
\Omega = NqT \; ,
\ee
free energy
\be
\label{58}
F = N(qT+\mu) \; ,
\ee
or for the Kramers potential
\be
\label{59}
K = - Nq \; .
\ee
As is seen from the latter relation, the quasipotential (25) is minus the
Kramers potential per particle. From the above equations, the pressure is
\be
\label{60}
p =-\; \frac{\Omega}{V} = -\rho q T \; ,
\ee
where $\rho$ is the mean density (51). The minimum of the quasipotential (25)
corresponds to that of the grand potential (67) or to the minimum of the free
energy (58), but to the maximum of the Kramers potential (59).

The quasipotential (25), due to the form of the effective Hamiltonian (26), can
be written as a sum
\be
\label{61}
q = \sum_\nu q_\nu \; , \qquad q_\nu \equiv -\; \frac{1}{N}\; \ln\;
{\rm Tr}_{\cH_\nu}\; e^{-\bt H_\nu} \; .
\ee
Thence, the grand potential (57) becomes
\be
\label{62}
\Omega =\sum_\nu \Omega_\nu \; , \qquad
\Omega_\nu \equiv - T\ln\; {\rm Tr}_{\cH_\nu} \; e^{-\bt H_\nu} \; .
\ee
And the free energy (58) also writes as a sum
\be
\label{63}
F = \sum_\nu F_\nu \; , \qquad F_\nu \equiv \Omega_\nu + \mu N_\nu \; .
\ee
However, one should not forget that the terms in the sums (61) to (63) depend
on thermodynamic parameters as well as on the phase probabilities, e.g.,
\be
\label{64}
\Omega_\nu =\Om_\nu(T,V,\mu,w_\nu) \; , \qquad
F_\nu = F_\nu(T,V,N,w_\nu) \; ,
\ee
where $w_\nu$ is a function of $T,V,\mu$ or $T,V,N$, respectively. These terms
correspond to phase replicas of a heterophase system, but not to pure phases.
The latter would be described by the potentials
\be
\label{65}
\Om_\nu'=\Om_\nu'(T,V_\nu,\mu) =\Om_\nu(T,V_\nu,\mu,1) \; , \qquad
F_\nu'=F_\nu'(T,V_\nu,N_\nu) = F_\nu(T,V_\nu,N_\nu,1) \; .
\ee
These do not compose the thermodynamic potentials (62) and (63) of a heterophase
system, that is,
\be
\label{66}
\Om\neq \sum_\nu \Om_\nu' \; , \qquad F \neq \sum_\nu F_\nu' \; .
\ee
The difference
\be
\label{67}
\Om_{int} \equiv \Om - \sum_\nu \Om_\nu' \; , \qquad
F_{int}\equiv F - \sum_\nu F_\nu'
\ee
between the total thermodynamic potentials (62) or (63) and the sums of the
related potentials (65), representing pure phases, can be associated with the
thermodynamic potentials of interphase layers [1]. Analogous conclusions concern
the Kramers potential (59), for which
\be
\label{68}
K = \sum_\nu \ln\; {\rm Tr}_{\cH_\nu}\; e^{-\bt H_\nu} \; .
\ee
But this potential is a function $K(\bt,V,\bt\mu)$ of not the most convenient
thermodynamic variables, because of which it is rarely employed in practice.

In this way, the whole heterophase system is characterized by a total
thermodynamic potential, like the grand potential (62) or free energy (63).
Such a potential writes as a sum of the terms representing phase replicas
of heterophase matter, so that these terms should not be confused with the
potentials of pure phases. A phase-transition point, defined in the theory
of heterophase fluctuations, can be rather different from a transition point
given by the conventional approach dealing solely with pure systems. In the
latter case, a phase transition between two phases, say the
melting-crystallization transition, is defined by the equalities
\be
\label{69}
\Om_1'(T,V,\mu) =\Om_2'(T,V,\mu) \; , \qquad
F_1'(T,V,N) = F_2'(T,V,N)
\ee
for the potentials of pure phases, which give, e.g., the transition
temperature $T'(\mu)$ or $T'(\rho)$, respectively. This temperature coincides
with the transition temperature $T_0$ in the theory of heterophase matter
only if $w_1\equiv 1$ and $w_2\equiv 0$ below $T_0$, while $w_1\equiv 0$,
$w_2\equiv 1$ above $T_0$. However, generally, $w_\nu\not\equiv 0,1$, and the
transition point in heterophase matter is different from that in pure-phase
systems. Thus, the phase transition temperature $T_0$ in heterophase matter
is defined as the point where the relation between $w_1$ and $w_2$ changes
so that
\be
\label{70}
w_1 > w_2 \quad (T=T_0-0) \; , \qquad w_1 < w_2 \quad (T=T_0+0) \; .
\ee
In addition, the existence of mesoscopic heterophase fluctuations may lead
to various pretransitional effects [1], which can be explained only taking
these fluctuations into account.

\section{Turbulent versus Chaotic Crystals}

The majority of crystals are associated with a periodic crystalline
structure. There exist as well quasicrystals, characterized by a quasiperiodic
spatial distribution of particles [39]. Crystals and quasicrystals can form
stable matter. The structure that is not uniform and neither periodic nor
quasiperiodic, can be called chaotic. Examples of solids with chaotic structure
are glasses and amorphous materials. In nature, such chaotic solids are
metastable. Ruelle [40] put forward a question asking if solids with random
spatial structure could form not metastable but stable states. Newell and
Pomeau [41] mentioned that such static chaotic spatial structures could exist
in principle, which was supported by a two-dimensional numerical modeling [42].
The solids with chaotic spatial structures can be named {\it chaotic crystals}.
Sometimes, they are also termed as turbulent, which, however, is clearly
a misnomer. Turbulence is a {\it spatio-temporal} phenomenon characterizing
unstable temporal evolution of chaotic spatial structures [43]. It is,
therefore, incorrect to call a static spatial structure turbulent. To dignify
a phenomenon by the name of turbulence compulsory implies the involvement
of chaotic temporal evolution. Therefore stable solids with a static random
spatial structure [40--42] can be correctly classified only as chaotic
crystals.

The existence of temporal heterophase fluctuations is what can render solids
really turbulent. Randomly fluctuating in time and chaotically distributed
in space mesoscopic heterophase fluctuations realize {\it heterophase
turbulence} [13]. A stable solid exhibiting heterophase turbulence is a
{\it turbulent crystal}. The averaging over heterophase configurations,
described is Section 3, leads to an effective quasiequilibrium picture of
heterophase turbulent matter, similarly to the averaged treatment of usual
turbulence [43].

{\it Heterophase turbulence} may develop in any condensed matter. Moreover,
it is heterophase turbulence which is the {\it dynamical cause of all phase
transitions}. The melting-crystallization transition, discussed in Section
4 is not an exception. Heterophase turbulence in a crystal shows up as
chaotically appearing and disappearing germs of liquid-like regions of
disordered lattice. The crystalline and fluid phases can be distinguished
between each other by their differing symmetries resulting in different
features of their spatial local densities (52) and (53). Competing phases
quite often differ in their symmetry, which, though, is not a mandatory
prerequisite for their distinction. Thermodynamic phases may, for instance,
having the same symmetry, possess different mean densities, as liquid and
gas. The possibility of heterophase turbulence in a mixture of two coexisting
phases with differing densities $\rho_1\neq\rho_2$ can be illustrated by
a lattice-gas model [44,45].

Let the phases be distinguished by their average densities $\rho_1$ and
$\rho_2$, so that
\be
\label{71}
\rho_1 >\rho_2 \; .
\ee
Suppose that the system consists of $N$ particles each of which can be
located in one of the $N_s$ sites of a crystalline lattice $\{\ba\}$, with
$i=1,2,\ldots,N_s$ and $N\leq N_s$. The field operator of a particle in
a $\nu$-phase can be presented as an expansion
\be
\label{72}
\psi_\nu(\br) = \sum_{in} c_{in\nu}\; e_{i\nu}\; \vp_n(\br-\ba_i)
\ee
over the Wannier functions $\vp_n(\br-\ba_i)$, labelled by a multi-index $n$,
with $e_{i\nu}=0,1$ being the occupation variable. Each site can be occupied
by not more than one particle, which implies the validity of the unipolarity
condition
\be
\label{73}
\sum_n c^\dgr_{in\nu}\; c_{in\nu} = 1 \; .
\ee
The particles are assumed to be well localized in the lattice sites, so that
the matrix elements over the Wannier functions $\vp_n(\br-\ba_i)$ for the
kinetic operator $\hat K\equiv -\hbar^2\nabla^2/2m_0$ are
\be
\label{74}
<mi|\hat K|nj> \; = \dlt_{mn}\; \dlt_{ij} K_0
\ee
and for the interaction potential $\Phi(\br-\br')$ are
\be
\label{75}
<mi,nj|\Phi|m'i',n'j'>\; = \dlt_{mn'}\; \dlt_{ij'}\;
\dlt_{nm'}\; \dlt_{ji'} \; \Phi_{ij} \; .
\ee
Then the replica Hamiltonian (46) reduces to
\be
\label{76}
H_\nu = w_\nu \sum_i (K_0 - \mu) e_{i\nu} +
\frac{1}{2}\; w_\nu^2 \sum_{i\neq j} \Phi_{ij} e_{i\nu} e_{j\nu} \; .
\ee
The average phase density (50) takes the form
\be
\label{77}
\rho_\nu = \frac{1}{V} \sum_i <e_{i\nu}> \; .
\ee
The chemical potential $\mu$ is defined by the normalization condition (51).

Introducing a quasispin operator
\be
\label{78}
S_{i\nu} \equiv e_{i\nu} -\; \frac{1}{2}
\ee
transforms the replica Hamiltonian (76) to the Ising-type representation
\be
\label{79}
H_\nu = E_\nu + \frac{1}{2}\; w_\nu^2 \sum_{i\neq j}
\Phi_{ij} S_{i\nu} S_{j\nu} - w_\nu \sum_i B_\nu S_{i\nu} \; ,
\ee
in which
$$
E_\nu = \frac{1}{2}\; N_s w_\nu \left ( K_0 - \mu + \frac{1}{4}\;
w_\nu \Phi\right ) \; ,
$$
\be
\label{80}
\Phi\equiv \frac{1}{N_s} \sum_{i\neq j} \Phi_{ij} \; , \qquad
B_\nu = \mu - K_0 - \; \frac{1}{2}\; w_\nu \Phi \; .
\ee
The phase density (77) becomes
\be
\label{81}
\rho_\nu \equiv \frac{N_s}{2V}\; (1 + m_\nu ) \; ,
\ee
where
\be
\label{82}
m_\nu \equiv \frac{2}{N_s} \sum_i < S_{i\nu}>
\ee
is an analog of a reduced magnetization. Condition (71), owing to Eq. (81),
yields
\be
\label{83}
m_1 > m_2 \qquad (\rho_1 >\rho_2 ) \; .
\ee
Normalization (51) can be rewritten as the equation
\be
\label{84}
n \equiv \frac{N}{N_s} = \frac{1}{2}\; w_1 (1 + m_1) + \frac{1}{2}\; w_2
(1 + m_2)
\ee
defining the chemical potential $\mu=\mu(n,T)$ as a function of the parameter
(84) and temperature $T$.

The considered two-density lattice model can correspond [13,45] to a stable
system whose thermodynamic potentials (62) or (63) are lower than the
potentials (65) of pure phases, so that
\be
\label{85}
\Om(T,V,\mu) < \Om_\nu'(T,V,\mu) \; , \qquad
F(T,V,N) < F_\nu'(T,V,N) \; .
\ee
This means that heterophase turbulence may stabilize condensed matter. That
is, turbulent crystals should be rather a rule than an exotic exception. Any
real solid can exhibit, under some conditions, especially in the vicinity of
transition points, heterophase turbulence, thus, becoming a turbulent crystal.

\section{Ferromagnet with Paramagnetic Fluctuations}

Heterophase fluctuations may arise in systems of different nature. Specific
features of the resulting heterophase matter can be well illustrated by
a model of ferromagnet with mesoscopic paramagnetic fluctuations [1].

The general form of the replica Hamiltonian of a lattice spin system reads
\be
\label{86}
H_\nu = w_\nu NK_\nu + \frac{1}{2}\; w_\nu^2 NU_\nu -
w_\nu^2 \sum_{i\neq j} J_{ij\nu}{\bf S}_{i\nu}\cdot{\bf S}_{j\nu} \; ,
\ee
where $N$ is the number of spins; $K_\nu$, kinetic parameter; $U_\nu$,
crystalline field; $J_{ij\nu}$, exchange interactions; ${\bf S}_{i\nu}$
are spin operators associated with a $\nu$-phase. Ferromagnetic and
paramagnetic phases can be distinguished by the order parameter
\be
\label{87}
\eta_\nu \equiv |< {\bf S}_{i\nu} >| \; ,
\ee
which, in the case of an ideal periodic lattice, does not depend on the
lattice-site index. Assume that $\nu=1$ labels the ferromagnetic phase, while
$\nu=2$, the paramagnetic phase. By this assumption,
\be
\label{88}
\eta_1\neq 1 \; , \qquad \eta_2 \equiv 0 \; .
\ee
The zero order parameter of the paramagnetic phase supposes that the case
without external magnetic fields is considered. In the presence of the latter,
the average magnetization of a paramagnet would be nonzero. Then, we could
distinguish the phases by the condition $\eta_1>\eta_2$, similarly to the
relations (71) or (83).

Keeping in mind long-range exchange interactions $J_{ij\nu}$, we shall employ
the mean-field approximation. The notation
\be
\label{89}
\eta\equiv \eta_1 \; , \qquad w = w_1 \; , \qquad J \equiv \frac{1}{N}\;
\sum_{i\neq j} J_{ij1}
\ee
will be used. Then for the probability of ferromagnetic phase, we find
\be
\label{90}
w = \frac{U_2 +K_2 - K_1}{U_1 + U_2 - 2J\eta^2} \; ,
\ee
which follows from Eq. (43) with
$$
\Phi_\nu = U_\nu - 2J_\nu \eta_\nu^2 \; , \qquad
J_\nu \equiv \frac{1}{N} \sum_{i\neq j} J_{ij\nu} \; .
$$
For the average spin (87), we have
\be
\label{91}
2\eta ={\rm tanh}\left ( w^2\; \frac{J}{T}\; \eta\right ) \; .
\ee
Here and in the following formulas, we shall set $k_B=1$. It is reasonable to
assume that the arising magnetic order does not influence much the crystalline
properties of matter, so that the kinetic terms $K_\nu$ and the crystalline
fields $U_\nu$ do not depend on the phase index $\nu$,
$$
K_1 = K_2 \; , \qquad U_1 = U_2 \equiv A \; .
$$
Then, with the notation
\be
\label{92}
u\equiv \frac{A}{J} \; ,
\ee
the ferromagnetic weight (90) reduces to
\be
\label{93}
w =\frac{u}{2(u-\eta^2)} \; .
\ee
Thus, instead of one equation for the average spin, as it would be in the
case of a pure phase, we have now two coupled equations, one for the average
spin (91) and another for the probability of ferromagnetic phase (93). Note
that the latter can be treated as an additional order parameter that becomes
necessary for correctly describing heterophase matter.

The analysis of the model of the heterophase ferromagnet, in the mean-field
approximation, shows that there exist two critical temperatures
\be
\label{94}
T_c = \frac{1}{8}\; J \; , \qquad T_c' =\frac{1}{2}\; J \; .
\ee
The overall behaviour of the system essentially depends on the value of
the dimensionless parameter (92), which measures the relative intensity of
disordering interaction $A$ to ordering interaction $J$. According to the
stability conditions (44) and (45), paramagnetic fluctuations in a ferromagnet
may exist only when $u>\eta^2>0$. Therefore, for some $u$ there can exist the
{\it nucleation temperature} $T_n$, defined by the condition
\begin{eqnarray}
\label{95}
\begin{array}{ll}
w\equiv 1 & (T\leq T_n) \; , \\
w< 1 & (T> T_n) \; .
\end{array}
\end{eqnarray}
Depending on the parameter $u$, we can observe five types of behaviour:

\vskip 2mm

(i) $u\leq 0$.

\vskip 1mm

Pure ferromagnetic phase, with $w\equiv 1$, is absolutely stable up to the
critical temperature $T_c'$, where the system becomes paramagnetic. Heterophase
ferromagnet can be only metastable, having the critical temperature $T_c$.

\vskip 2mm

(ii) $0 < u < \frac{3}{2}$.

\vskip 1mm

In the region of temperatures $0\leq T\leq T_n$, the system is a pure ferromagnet,
after which paramagnetic fluctuations appear, starting from the nucleation point
$T_n$, for $T>T_n$. The phase transition ferromagnet-paramagnet becomes
a first-order transition, occurring at the point $T_0$, for which $T_c <T_0<T_c'$.

\vskip 2mm

(iii) $0 < u < \frac{1}{2}$.

\vskip 1mm

The nucleation point $T_n$ exists only for this interval of $u$, such that
$0<T_n<T_c'$. When $u\ra 0$, then $T_n\ra T_c'$ and if $u\ra 0.5$, then
$T_n\ra 0$. For  $u>0.5$, there is no nucleation point. The first-order phase
transition takes place at $T_0$.

\vskip 2mm

(iv) $u =\frac{3}{2}$.

\vskip 1mm

Ferromagnet is always heterophase. The transition temperature $T_c$ becomes
a tricritical point, where critical indices change by a jump.

\vskip 2mm

(v) $u>\frac{3}{2}$.

\vskip 1mm

Heterophase ferromagnet is stable. The second-order transition happens at the
critical temperature $T_c$.

\vskip 2mm

The behaviour of the order parameter, ferromagnetic phase probability, and
other thermodynamic characteristics is illustrated in the following figures.
The order parameter $\eta_1$, defined as the average spin (87), for different
values of $u$, is shown in Fig. 4 as a function of $\Theta/J$, with
$\Theta=k_BT$.

The probability of ferromagnetic phase $w_1\equiv w$, satisfying Eq. (93), is
presented in Fig. 5 as a function of the reduced temperature $\Theta/J$.

The average magnetization
\be
\label{96}
M \equiv w \eta
\ee
as a function of $\Theta/J$ is depicted in Fig. 6. For the values $u$ in the
range $-0.25<u<0$, there exists a maximum of the magnetization $M_m$ at the
temperature $T_m$, where
\be
\label{97}
M_m = \frac{\sqrt{-u}}{4} \; , \qquad
T_m = \frac{\sqrt{-u}\; J}{16{\rm arctanh}(2\sqrt{-u})} \; .
\ee
These quantities are shown in Fig. 7 as functions of $u\equiv A/J$.

The first-order phase transition to the paramagnetic state occurs at temperature,
where the free energy per particle
$$
f \equiv \frac{F}{N} = -\; \frac{T}{N}\; \ln\; {\rm Tr}\; e^{-\tilde H/T}
$$
for the heterophase ferromagnet, given by the expression
\be
\label{98}
\frac{f}{J} =\left ( w^2 - w +\frac{1}{2}\right ) u + w^2\eta^2 -\;
\frac{T}{J}\; \ln\left [ 4{\rm cosh}\left ( w^2\;
\frac{J}{T}\; \eta\right ) \right ] \; ,
\ee
coincides with the value
\be
\label{99}
\frac{f_0}{J} = \frac{u}{4}\; - \; \frac{T}{J}\; \ln 4 \; .\,
\ee
corresponding to $\eta=0$ and $w=0.5$. Functions (98) and (99) are illustrated
in Fig. 8 for $u=0.6$.

The relative entropy
\be
\label{100}
\sigma \equiv \frac{E-F}{NT} \qquad (E \equiv\; <\tilde H>)
\ee
takes the form
\be
\label{101}
\sigma = - 2w^2 \; \frac{J}{T}\; \eta^2 +\ln\left [ 4{\rm cosh}\left ( w^2\;
\frac{J}{T}\; \eta\right ) \right ] \; ,
\ee
which is depicted in Fig. 9 for different $u$.

The specific heat
\be
\label{102}
C_H \equiv \frac{1}{N}\; \frac{\partial E}{\partial T} =
T\; \frac{\partial\sigma}{\partial T}
\ee
in portrayed in Fig. 10, from where it is seen that metastable heterophase
ferromagnets, with $u<0$, possess specific-heat maxima below the critical
temperature.

When the critical point $T_c$ becomes tricritical at $u=3/2$, the critical
indices change by a jump. Thus, for the critical indices $\al$ and $\bt$,
characterizing the asymptotic behaviour of the specific heat and magnetization
as
\be
\label{103}
C_H \sim (-\tau)^{-\al} \; , \qquad M\sim (-\tau)^\bt \; ,
\ee
where
$$
\tau \equiv \frac{T-T_c}{T_c} \ra - 0 \; ,
$$
we obtain
\begin{eqnarray}
\label{104}
\al=\left\{ \begin{array}{ll}
0\; , & u\neq 3/2 \\
1/2 \; , & u = 3/2 \; ,
\end{array}\right.
\bt =\left\{ \begin{array}{ll}
1/2\; , & u\neq 3/2 \\
1/4\; , & u =3/2\; .
\end{array}\right.
\end{eqnarray}
In any case, $\al+2\bt=1$. If one calculates the critical indices with $\tau$
tending to a small but finite value, instead of $\tau\ra-0$, then one gets the
effective
indices
\be
\label{105}
\al_{eff} = -\; \frac{\ln C_H}{\ln(-\tau)} \; , \qquad \bt_{eff} =
\frac{\ln M}{\ln(-\tau)} \; .
\ee
These are pictured in Fig. 11 for $|\tau|=10^{-8}$.

Finally, Fig. 12 demonstrates the temperature variation of the average spin,
defined by Eq. (91), for different values of the parameter (92). Only the
solutions for stable states are presented.

\section{Conclusion}

Mesoscopic heterophase fluctuations may arise in any kind of condensed matter.
This is illustrated here for crystals with liquid-like fluctuations, for a
two-density lattice mixture, and for ferromagnets with paramagnetic fluctuations.
Of course, these fluctuations appear not always, but their existence depends on
the system parameters as well as on thermodynamic variables. The presence of
such fluctuations can make the system thermodynamically more stable. Their
appearance may provoke different pretransitional phenomena and transitional
anomalies.

The fluctuations are mesoscopic in size, since they represent the germs of
competing thermodynamic phases, which requires that the characteristic length
of such a germ be much larger than the mean interparticle distance. At the same
time, this characteristic length can be much smaller than the size of the
whole sample. If the fluctuations are not static, but evolve in time, appearing
and disappearing, their typical lifetimes are to be essentially longer than the
local-equilibrium time, but can be much shorter than the observation time, which
means that such temporal fluctuations are mesoscopic in time. In any case, these
fluctuations are randomly distributed in space.

The spatial randomness of the mesoscopic fluctuations makes it possible to
realize the averaging over heterophase configurations and to define an effective
renormalized Hamiltonian. In the course of this averaging, novel quantities
emerge, representing the probabilities of thermodynamic phases. These
probabilities are defined as minimizers of the related thermodynamic potentials.

The theory of heterophase fluctuations [1] provides a general approach for
treating a wide class of materials whose properties can be strongly influenced
by the existence of such fluctuations. At the same time, many phenomena occurring
in condensed matter, which look anomalous without taking account of the
mesoscopic fluctuations, can be explained by the influence of the latter.

\newpage

\begin{center}
{\large{\bf Figure Captions}}
\end{center}

\vskip 2cm

{\bf Fig. 1}. The crystalline phase probability as a function of temperature.

\vskip 5mm

{\bf Fig. 2}. The liquid phase probability as a function of temperature.

\vskip 5mm

{\bf Fig. 3}. The glassy phase probability versus temperature.

\vskip 5mm

{\bf Fig. 4}. The order parameter $\eta_1\equiv\eta$, given by Eq. (91),
for different parameters (92): $u=0.5$ (curve 1), $u=0.6$ (curve 2),
$u=0.8$ (curve 3), $u=1.5$ (curve 4), $u=\infty$ (curve 5), $u=-1$ (curve 6),
$u=-0.25$ (curve 7), $u=-0.1$ (curve 8), $u=-0.01$ (curve 9),
$u=-0.001$ (curve 10).

\vskip 5mm

{\bf Fig. 5}. The ferromagnetic phase probability $w_1\equiv w$, defined by
Eq. (93), as a function of relative temperature, for the same values of $u$
as in Fig. 4.

\vskip 5mm

{\bf Fig. 6}. The average magnetization (96) as a function of relative
temperature for different values of $u$ as in Fig. 4.

\vskip 5mm

{\bf Fig. 7}. The maximum of magnetization $M_m$ and the related temperature,
as defined in Eq. (97), versus $u\equiv A/J$.

\vskip 5mm

{\bf Fig. 8}. Free energies per particle, given by Eqs. (98) and (99)
for $u=0.6$.

\vskip 5mm

{\bf Fig. 9}. Relative entropy (101) for the values of $u$ as in Fig. 4.

\vskip 5mm

{\bf Fig. 10}. Specific heat (102) for different $u$: $u=0.6$ (curve 2),
$u=0.8$ (curve 3), $u=1.5$ (curve 4), $u=3$ (curve with no number),
$u=\infty$ (curve 5), $u=-1$ (curve 6).

\vskip 5mm

{\bf Fig. 11}. Effective critical indices (105) for $\tau=-10^{-8}$, versus
$u\equiv A/J$.

\vskip 5mm

{\bf Fig. 12}. Average spin, satisfying Eq. (91), for different values of $u$,
which are denoted by the numbers close to the related curves.

\end{document}